\documentclass[12pt]{iopart}
\usepackage{iopams}  
\usepackage{graphicx}
\usepackage{color}

\newcommand{\mysection}[1]{\section{#1}}

\begin{document}

\title[Inelastic chaotic scattering on a Bose-Einstein condensate]{Inelastic chaotic scattering on a Bose-Einstein condensate}

\author{Stefan Hunn$^1$, Moritz Hiller$^{1,2}$, Doron Cohen$^3$, Tsampikos Kottos$^4$, and Andreas Buchleitner$^1$ }

\address{
  $^1$ Physikalisches Institut, Albert-Ludwigs-Universit{\"a}t Freiburg, Hermann-Herder-Str. 3, 79104 Freiburg, Germany\\
  $^2$ Institute for Theoretical Physics, Vienna University of Technology, Wiedner Hauptstra{\ss}e 8-10/136, A-1040 Vienna, Austria\\
  $^3$ Department of Physics, Ben-Gurion University of the Negev, Beer-Sheva 84105, Israel \\
  $^4$ Department of Physics, Wesleyan University, Middletown, Connecticut 06459, USA and 
  MPI for Dynamics and Self-Organization, Am Fa{\ss}berg 17, D-37077 G{\"o}ttingen, Germany}

\begin{abstract}
We devise a microscopic scattering approach to probe the excitation spectrum of a Bose-Einstein condensate.
We show that the experimentally accessible scattering cross section exhibits universal Ericson fluctuations,
with characteristic properties intimately related to the underlying classical field equations.
\end{abstract}

\maketitle

\mysection{Introduction}
Bose-Einstein condensates (BEC) in optical lattices provide a versatile tool to address experimentally 
a variety of questions that emerge in diverse fields ranging from quantum information and many-body quantum phase
transitions to solid-state transport and atomtronics. An important element of these studies is the development and
implementation of methods which allow for an accurate measurement of the properties of the condensate. Among the
most popular ones are time-of-flight and Bragg-spectroscopy \cite{SICSPK99, SMSKE04,MO06}
techniques which result in the destruction of the BEC, whereas only few works consider a scattering setup that leaves the
condensate intact \cite{MMR07b}. Specifically, the main focus of the present literature is on photon-atom scattering, while only
very recently matter-wave scattering was experimentally implemented to infer the spatial ordering of ultracold atomic crystals \cite{GPRS11}, and
theoretically proposed to probe the Mott insulator to superfluid transition of the condensate ground state \cite{florian}.
However, excited states 
are naturally populated in experiments which
probe non-trivial BEC dynamics \cite{AGFHO05}.
The rapidly emerging complexity of the many-body dynamics -- manifest, e.g., in dynamical instabilities \cite{MO06} -- is a direct
fingerprint of the complex underlying spectral structure, which is itself reflected in the -- in general chaotic -- classical limit of the
Bose-Hubbard model (achieved in the limit of large particle numbers). It is therefore timely to explore possible experimental strategies
to probe these spectral features, in a non-destructive manner.
In our present contribution, we show how an inelastically scattered probe particle can reveal the state of a BEC target in the parameter
regime of spectral chaos. Due to the inherent sensitivity of spectral cross sections under such conditions, a robust
characterization requires a statistical approach, which can be further sharpened by semiclassical considerations.

\begin{figure}[t]
\begin{center}
  \includegraphics[width=.72\columnwidth,keepaspectratio]{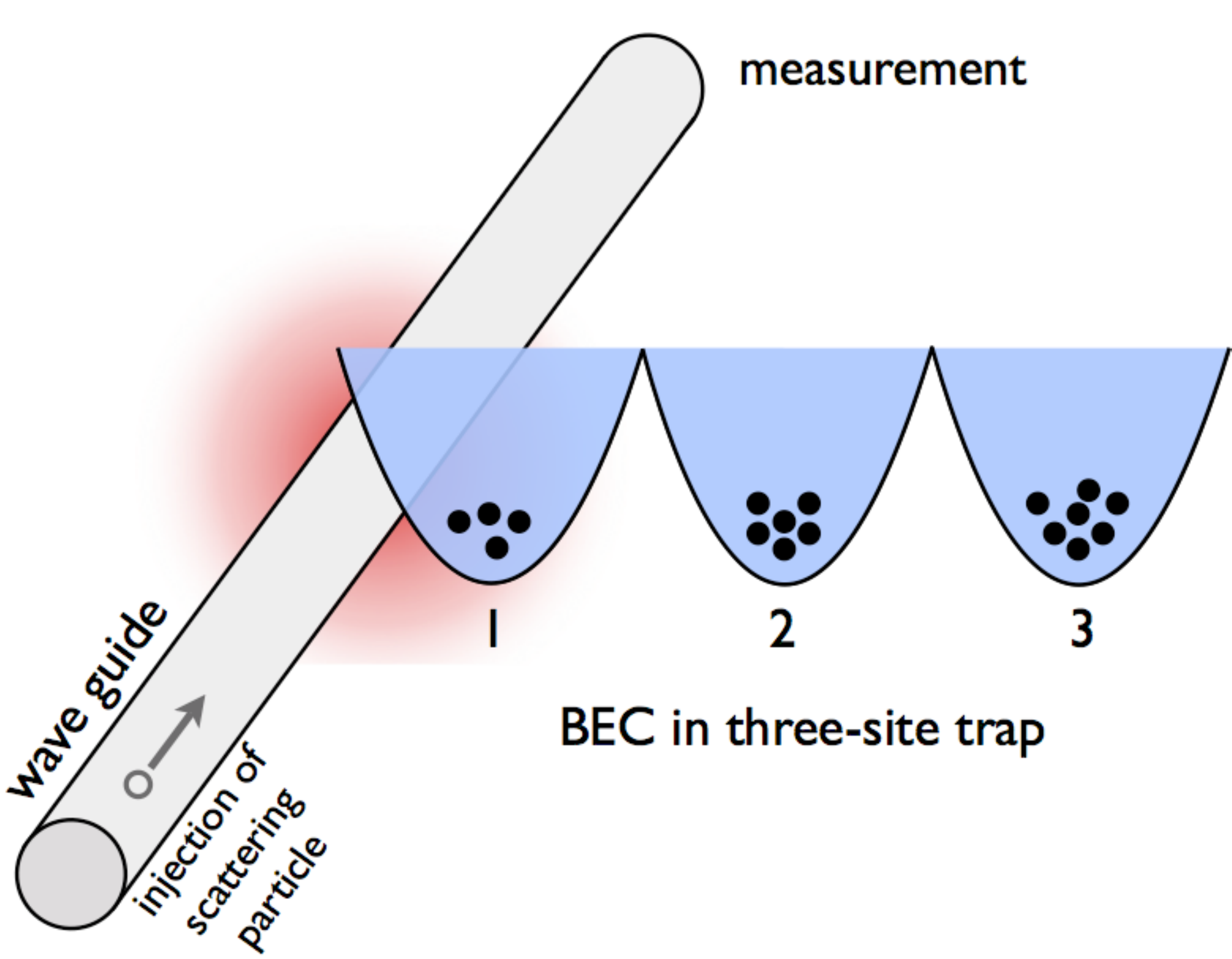}
\end{center}
\caption{(color online). Scattering setup: 
The probe particle is injected into a wave guide, and locally exchanges energy with a BEC
confined by a three-site optical potential, in the contact region between wave guide and site one
of the lattice. The inelastic scattering cross section measured on
the wave guide's exit bears direct information on the state of the BEC.
} 
\label{fig.setup}
\end{figure}

%
\mysection{Model}
The scattering setup that we have in mind is shown in Fig.~\ref{fig.setup}: A probe particle with momentum $k$ moves 
in a waveguide which is placed in the proximity of a BEC confined by an optical lattice. When the particle approaches
the condensate, it interacts with the latter -- much as the condensate particles between themselves -- leading to an
exchange of energy. The particle energy on exit from the waveguide defines the scattering cross section.
The dynamics of the process is generated by 
\begin{equation}
H_{\rm tot} = H_{\rm TB} \otimes {\hat 1} + {\hat 1} \otimes H_{\rm BH} + H_{\rm int} \, ,
\label{eq.H_tot}
\end{equation}
where ${\hat 1}$ denotes the identity operator. In (\ref{eq.H_tot}), $H_{\rm BH}$ represents the BEC target's  
Bose-Hubbard-Hamiltonian \cite{JBCGZ98}
\begin{equation}
H_{\rm BH} = 
\frac{U}{2}\sum_{i=1}^L \hat{n}_i(\hat{n}_i - 1) -  K  \sum_i \left[ \hat{b}_i^{\dagger} \hat{b}_{i+1}\, +{\rm h.c.} \right] 
\label{eq1_BHH}
\end{equation}
of $N$ interacting bosons on an $L$-site optical lattice, with
$\hat{b}_i^{(\dagger)}$ the bosonic annihilation (creation) operators, and $\hat{n}_i=\hat{b}_i^\dagger\hat b_i$ 
the particle number at site $i$. $U$ and $K$ parameterize the on-site interaction and the inter-site
tunneling strength, respectively. In the macroscopic limit $N\rightarrow\infty$ ($U N$ fixed), the dynamics of the condensate
is well described by mean-field theory, i.e. the discrete Gross-Pitaevskii equation. In this limit, the quantum operators $b_i^{(\dagger)}$
are replaced by $L$ complex amplitudes $A_i^{(*)}$ of a single-particle field. The Hamiltonian (\ref{eq1_BHH}) then reads
\begin{equation}
\frac{{\cal H}_{\rm GP}}{N}  = \frac{UN}{2} \sum_{i=1}^L |A_i|^4  - K  \sum_{i} \left[A_i^* A_{i+1} + {\rm  c.c.} \right]  \, ,
\label{eq.GP}
\end{equation}
where the $A_i$ are time-dependent, and obey the canonical equations $i\partial A_i/\partial t=\partial{\cal H}_{\rm GP}/\partial A_i^*$.

The waveguide in our scattering scheme is (for mathematical convenience) modeled by two semi-infinite tight-binding (TB) leads with hopping term $J$ and 
lattice spacing $a=1$. 
The particle's energy in the momentum eigenstate $|k_m\rangle$ of each lead is $\epsilon_m 
=-2 J \cos(k_m)$, with corresponding velocity $v_m = 2 J \sin(k_m)$ \cite{Datta95}.
These two leads are coupled with strength $J_0$ to the central site $j=0$, which is closest to the BEC.
$J_0$ thus controls the effective coupling of the projec\-tile-target interaction region to the asymptotically free states of the lead.
The projectile Hamiltonian is: 
\begin{equation}
\label{Hwaveguide} 
H_{\rm TB} = \Big[ -J \sum_{j\ne -1,0} \hat{c}_j \hat{c}_{j+1}^{\dagger}
-  J_0 \sum_{j=-1,0} \hat{c}_j \hat{c}_{j+1}^{\dagger} \Big] + {\rm h.c.} \, ,
\end{equation}
with $\hat{c}_j^{(\dagger)}$ the annihilation (creation) operators of the probe particle 
at site $j$ of the TB lead.

Finally, the probe-target interaction $H_{\rm int}$ is assumed to be of similar type (i.e. short range) as the bosonic
inter-particle interaction in the condensate:
\begin{equation}
\label{eq.H_int}
H_{\rm int} =  \alpha \ \hat{c}^{\dagger}_0 \hat{c}_0 \otimes \hat{n}_1\, .
\end{equation}
For non-vanishing tunneling coupling $K$, $H_{\rm int}$ induces transitions between different eigenmodes of the condensate,
what renders the scattering process inelastic. In this sense, $\alpha\!>\!0$ controls the inelasticity.
As the setup does not allow particle exchange, target bosons and probe particle are distinguishable at all times.
Thus, the latter can be chosen at the convenience of the experimentalist to implement the interaction described by (\ref{eq.H_int}).
In the macroscopic limit, the interaction Hamiltonian becomes time dependent, and is given by 
${\cal H}_{\rm int}/N =  \alpha |A_1|^2 \ c_0^{\dagger}c_0 $.

\mysection{Scattering matrix} 
Given the total Hamiltonian (\ref{eq.H_tot}) and the asymptotic freedom of the probe particle, we can now define the scattering
matrix of our problem, as the fundamental building block for our subsequent observations:
For the BEC initially prepared in an energy eigen\-state $|E_m\rangle$, and the probe particle injected with an energy 
$\epsilon_m$, the total system energy is ${\cal E}=E_m + \epsilon_m$. %
The open channels (modes) of the scattering process are then determined by energy conservation and 
characterized by the kinetic energy $\epsilon_{n}={\cal E}-E_{n}$ of the outgoing probe particle.
The transmission block of the scattering matrix can be derived from the Green's function of a particle at site $j=0$, with two
semi-infinite leads attached, and reads \footnote{The derivation of this formula in the present work is a tight-binding generalization of the continuum limit that has been analyzed in  \cite{BC08}. Details on the derivation can be found in \cite{Hunn09}.}:
\begin{equation}
[{\hat S}_T]({\cal E}) =\sqrt{\hat v} \frac{i\,\gamma}{(1-\gamma)  [{\cal E}- {\hat H}_{BH}] - \alpha {\hat n}_1 + i\gamma \hat{v} } 
\sqrt{\hat v} \, ,
\label{eq.S_T}
\end{equation}
where $\gamma \equiv(J_0/J)^2$, and ${\hat v}$ is the velocity operator. In the eigenbasis of the BH Hamiltonian, 
 $[{H}_{BH}]_{nm}=E_n \delta_{nm}$ and ${v}_{nm}=v_n \delta_{nm}$ are diagonal matrices, while $Q_{nm}\equiv \langle E_n| \hat{n}_1 | E_m\rangle$ is not.
For $\gamma=1$, Eq.~(\ref{eq.S_T}) coincides with the $S$-matrix for inelastic electronic scattering 
derived in \cite{BC08}.
In our setup, $\gamma<1$ can be regarded as a potential barrier that reduces the coupling between the leads and the scattering region
(i.e. for $\gamma=0$ the latter is isolated and the probe particle is perfectly reflected).
As $\gamma$ is increased from zero, one observes a crossover from a regime of well-resolved, narrow resonances to a regime of overlapping resonances.
Whereas the former regime would in principle allow us to directly infer the BH spectrum, it puts very hard demands on the experimental precision as far as preparation and measurement are concerned.
Instead, we find that in the latter regime one obtains an experimentally robust, fluctuating scattering signal that bears information
on the target's spectrum (around $\gamma=0.001$). This information is lost when $\gamma$ becomes very large compared to the
mean resonance spacing such that these characteristic fluctuations are washed out.

\begin{figure}[t]
\centerline{\includegraphics[width=0.8\columnwidth,keepaspectratio]{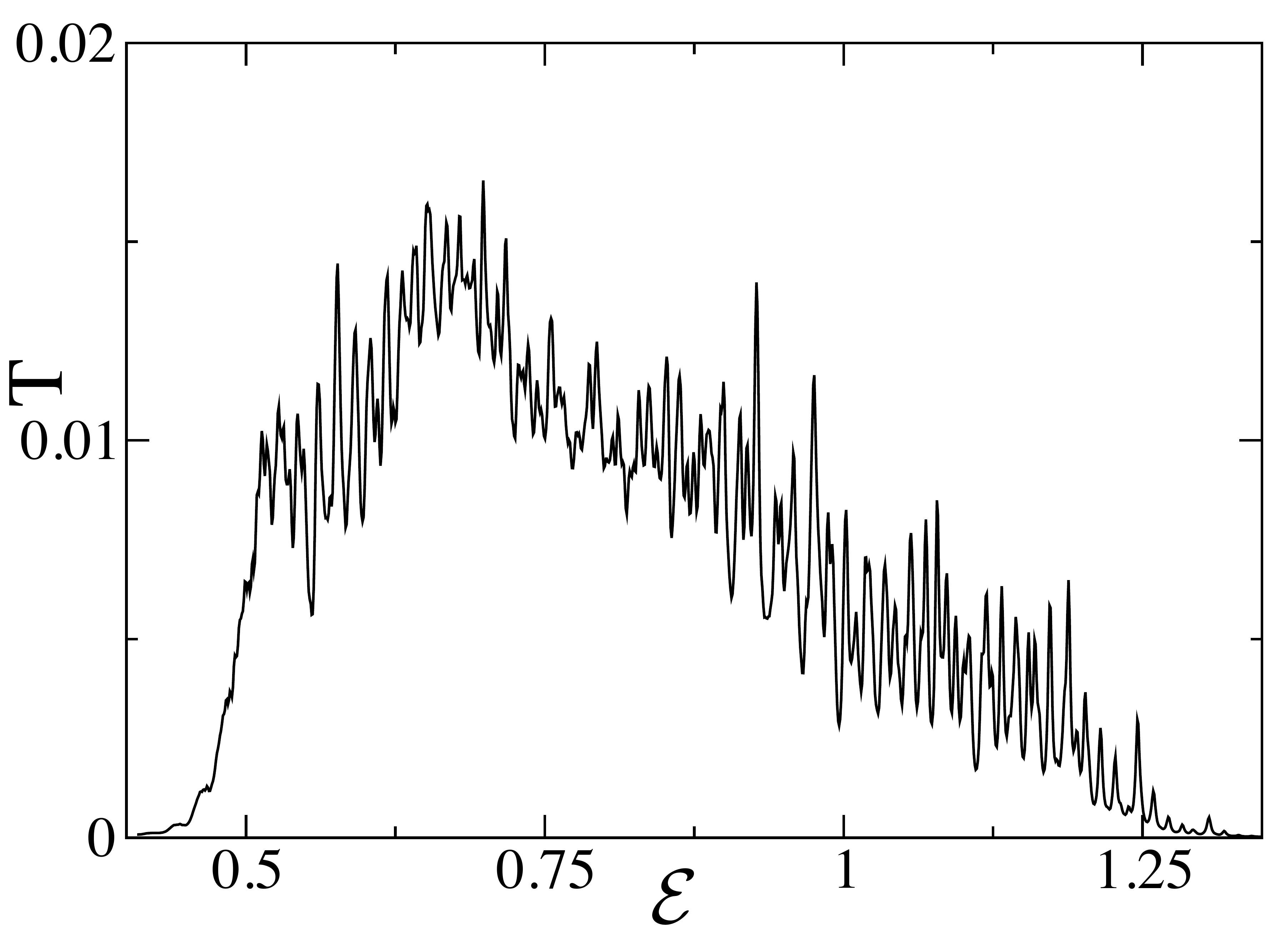}}
\caption{(color online).  
Transmission $T_{m}$, averaged over the channels $m=400-430$ in the chaotic regime, versus the total energy ${\cal E}$ in units of $K$, 
for $\gamma=0.001$, $N=38$ bosons, and $\alpha=5.0 K$. 
} 
\label{fig.transmission}
\end{figure}

%
\mysection{Chaotic scattering}
In our present contribution, we will investigate the properties of a probe particle scattering on a BEC 
that is described by the Hamiltonian (\ref{eq1_BHH}).
In contrast to Ref.~\cite{florian}, we are not concerned with the well-understood ground-state properties and the associated superfluid to Mott-insulator transition, which takes place at rather large interaction strengths $U$.
Instead, we focus on the complex properties of excited states at intermediate interaction strengths.

Namely, for $L>2$ and values ($3 \lesssim u  \lesssim 12$) of the control parameter $u=UN/2 K$, the classical Hamiltonian ${\cal H}_{GP}$ (\ref{eq.GP}) generates chaotic dynamics \cite{HKG09,LFYL07,WYFCL06,CFFCSS90, TWK09}.
Quantum manifestations thereof were investigated in a series of publications with emphasis on the statistical properties of the energy spectra \cite{BES90,FP03,BK03}. 
For a two-site lattice ($L=2$) \cite{SOD11}, inelastic scattering revealed immediate fingerprints of the fully integrable mean-field dynamics.
Here, we consider a three-site BH Hamiltonian ($L=3$), intermediate values of the control parameter around $u=5$,
what corresponds to the maximally chaotic regime \cite{HKG09}, large filling factors of the lattice (i.e. $N= 38$ bosons), and a condensate that is initially prepared in an energy eigenstate $|E_m\rangle$ in the bulk of the BH spectrum, where the target dynamics is predominantly chaotic.
The presence of chaos {\em qualitatively} alters the resulting physics:
In stark contrast to the integrable case $L=2$, the scattering quantities will strongly fluctuate (depending, e.g., on the probe particle's energy),
what requires a statistical analysis of the scattering signal.
On the other hand, chaos is expected to yield universal behavior, i.e., the obtained results should not depend on the details of the target system.
As for the other scattering parameters, we set $J=E_{max}- E_{min}=211K$, equal to the (numerically inferred) spectral width of the BH Hamiltonian,
such that all scattering channels are open, and fix $\gamma=0.001$.

\begin{figure}[t]
\begin{center}
  \includegraphics[width=\columnwidth,keepaspectratio]{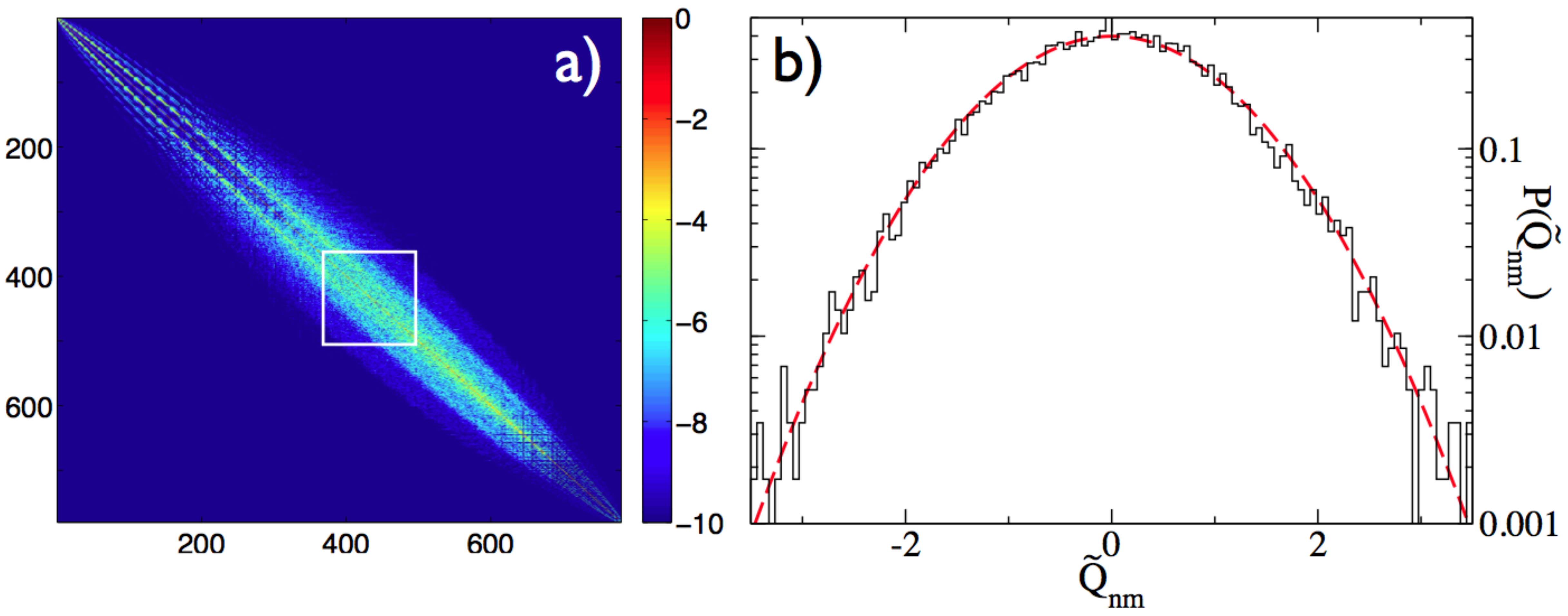}
\end{center}
\caption{(color online). 
{\it Left:} Logarithmically color-coded snapshot of the underlying interaction matrix $Q_{nm}$, for $u=5$.
{\it Right:} Distribution $P(\tilde Q_{nm})$ of the unfolded off-diagonal interaction matrix elements  $\tilde Q_{nm}$ taken from a sub-matrix in the bulk of $Q_{nm}$ (indicated by the white box in the left panel; the corresponding indices run from $n=m=360$ to $n=m=500$).
The distribution $P(\tilde Q_{nm})$ nicely resembles the dashed red line which represents the Gaussian distribution with unit variance $N(0,1)$.}
\label{fig.Q_prop}
\end{figure}

How do the chaotic spectral properties of the BEC manifest in a scattering experiment as sketched in Fig.~\ref{fig.setup}? 
The experimentally most easily accessible observable is 
the transmission $T_m({\cal E}) = \sum_n |[S_{T}]_{nm}|^2$. 
It denotes the probability that a probe particle with incoming energy $\epsilon_m$ exits the scattering area in any of
the outgoing channels  $\epsilon_n$. In Fig.~\ref{fig.transmission}, we show $T_m$ versus the total energy ${\cal E}$, averaged over
30 incoming channels $\epsilon_m$ in the middle of the spectrum, i.e. in the chaotic regime, where approximately $180$
outgoing channels contribute to $T_m$:

Strong fluctuations dominate the transmission signal -- a first indicator for the complexity of the target. 
In order to understand how this complexity enters in the scattering process, we turn to the scattering matrix (\ref{eq.S_T}).
In the $\{|E_n\rangle\}$ basis, in which the latter is evaluated, the interaction operator $\hat{n}_1$ becomes the only non-diagonal quantity on the r.h.s. of (\ref{eq.S_T}), and thus gives rise to inelastic scattering processes.
A closer inspection of the corresponding matrix $Q_{nm} \equiv \langle E_n| \hat{n}_1 | E_m\rangle$
(see Fig.~\ref{fig.Q_prop}a)) shows that for intermediate energies corresponding to the center of the matrix (i.e., in the chaotic energy regime) the matrix is
banded and its off-diagonal elements look rather erratically distributed. 
We analyze these elements, taken from the box indicated in Fig.~\ref{fig.Q_prop}a), and consider their distribution $P(\tilde Q_{nm})$.
Here, $\tilde Q_{nm}=Q_{nm}/\sqrt{\langle|Q_{nm}|^2\rangle}$ represent the off-diagonal matrix elements which are rescaled by their local standard deviation, the latter being calculated in a small, moving energy window.
This ``unfolding" is necessary to remove system-specific features and reveal universal properties of the interaction-operator $\hat n_1$ (see, e.g. \cite{HKG09}).
The resulting distribution is shown in Fig.\ref{fig.Q_prop}b) together with a Gaussian of unit variance.
The good agreement with the latter indicates that (neglecting higher-order correlations) the matrix elements are essentially independent random numbers, in perfect agreement with the predictions for quantum systems that possess a chaotic classical counterpart \cite{Berr77,AW92,PR93}.
We conclude that $S_T$ and thus all scattering quantities inherit their complexity from $Q$, since the latter represents the key ingredient in the scattering matrix.


\begin{figure*}[t]
\begin{center}
    \includegraphics[width=\textwidth,height=6cm]{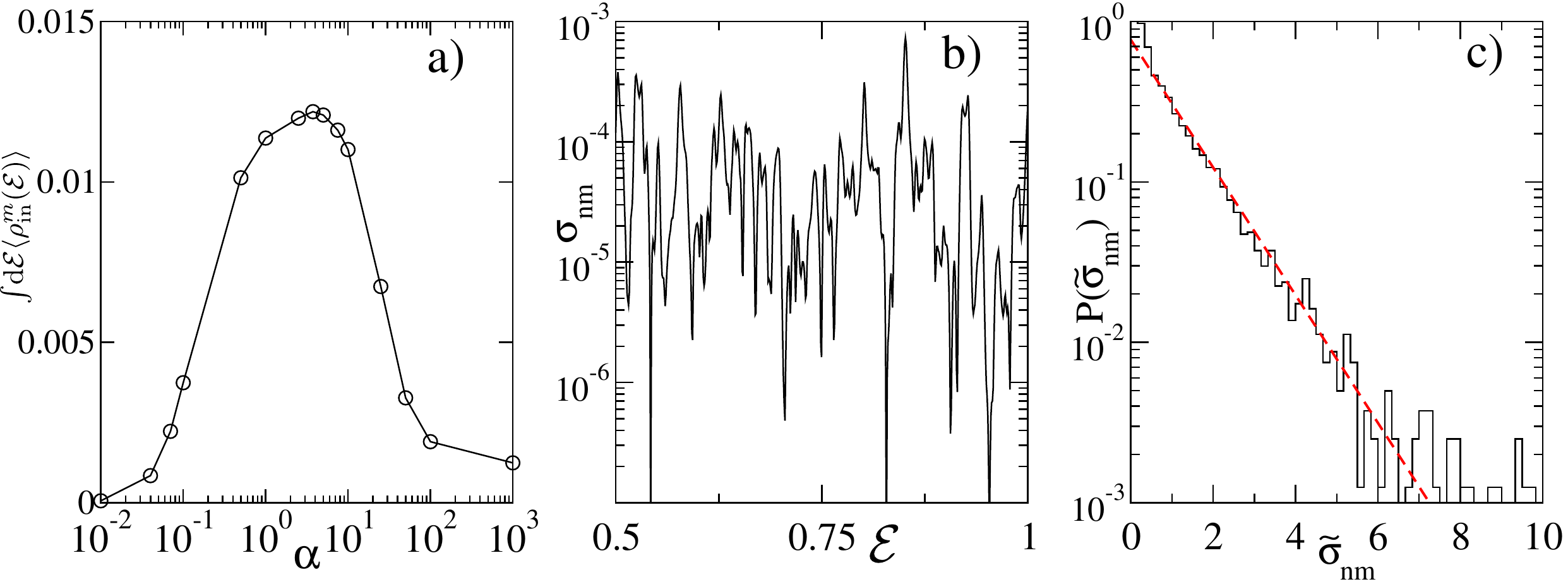}
\end{center}
\caption{(color online). 
{\it Left:}
Integrated total inelastic cross section $\int {\rm d}{\cal E} \langle\rho^m_{\rm in}({\cal E})\rangle$, for the same parameters as in
Fig.~\ref{fig.transmission}, averaged over the channels $m = 400 - 430$ in the chaotic regime, versus the inelasticity parameter
$\alpha$ in units of $K$. The integration runs over the entire energy axis.
{\it Middle:} 
A representative inelastic cross section $\sigma_{nm}$ versus the total energy ${\cal E}$ in units of $J$, for the same parameters and
$\alpha=5.0 K$.
{\it Right:} Histogram of the normalized inelastic cross section $P(\tilde\sigma_{nm})$, for
fifteen different channels $\sigma_{400m}$ ($m=401-415$) in the chaotic regime and identical parameter values.
The data perfectly match the dashed straight line exponential fit.
} 
\label{fig.rho_in}
\end{figure*}

To gain insight in the role of the parameter $\alpha$ that controls the inelasticity induced by $Q$, we next consider
the total inelastic scattering cross section
\begin{equation}
\label{elastic}
\rho^m_{\rm in}({\cal E}) = 2\sum_{n\ne m}|[S_T({\cal E})]_{nm}|^2   \, ,
\end{equation}
which essentially resembles $T_m$, except for the direct processes. 
For a given value of $\alpha$, we integrate over the energy axis, to obtain robust results, unaffected by the sensitive energy
dependence of  $\rho^m_{\rm in}({\cal E})$.
Fig.~\ref{fig.rho_in}a) shows that 
$\int {\rm d}{\cal E}\langle \rho^m_{\rm in}({\cal E}) \rangle_m$ takes its maximal value for intermediate values of the
inelasticity parameter $\alpha$, while it vanishes in the limit of small and large $\alpha$.
In the former case, the probe particle is directly transmitted, since (\ref{eq.S_T}) with $\alpha\approx0$ becomes diagonal,
while it is directly reflected in the latter case - as evident from (\ref{eq.S_T}) with $\alpha\gg1 K$ in the denominator. 
Consequently, only for intermediate $\alpha$-values around $\alpha\sim5 K$ can we expect to infer information on the condensate from the probe particle's exit energy.

\mysection{Ericson fluctuations}
Beyond total cross sections there is nontrivial dynamical information encoded in the {\em partial} inelastic cross
sections $ \sigma_{nm}({\cal E}) \equiv |[S_T({\cal E})]_{nm}|^2$, which quantify the probability 
for a transition from a state $E_m$ to a state $E_n$ of the target (or, equivalently, from an energy $\epsilon_m$ 
to an energy $\epsilon_n$ of the projectile).     
In Fig.~\ref{fig.rho_in}b) we show $ \sigma_{nm}({\cal E}) $, for the same parameter values as the transmission in
Fig.~\ref{fig.transmission}. 
We observe much stronger fluctuations than for the total transmission, what is simply due to the fact that the latter
implies an additional effective averaging over many scattering channels. As we will show now, this sensitive dependence
on the energy is an unambiguous trait of (universal) Ericson fluctuations, that were hitherto only reported in the context of
nuclear \cite{EM66} and atomic physics \cite{SW05}, as well as in microwave experiments \cite{SGSS03},
and are also connected to universal conductance fluctuations in mesoscopic physics \cite{Weid90,Sora09}.

The rapid fluctuations of the cross section are due to interference effects between overlapping resonances:
The scattering amplitudes $[S_T]_{nm}$ can be represented by a many-resonance
Breit-Wigner formula, where each individual term in the sum is 
a random variable whose stochastic properties originate from the statistically independent Gaussian distributed elements of the interaction matrix.
Then, due to the central limit theorem, one expects that real and imaginary part of the scattering matrix elements
are Gaussian distributed random numbers with zero mean.
In other words, the Gaussian distribution of the interaction matrix elements gives rise to Gaussian distributed real and imaginary part of $[S_T]_{nm}$.
This results in an exponential distribution \cite{EM66} $P(\tilde\sigma_{nm})=\exp[-\tilde\sigma_{nm}]$ of the normalized
inelastic cross section $\tilde\sigma_{nm}=\sigma_{nm}/\bar\sigma_{nm}$,
where $\bar\sigma_{nm}$ denotes the average inelastic cross section in the energy interval $\Delta {\cal E}$
(that is small compared to classical energy scales). This expectation is clearly confirmed by
our numerical data presented in Fig.~\ref{fig.rho_in}c).

The central figure of merit to identify Ericson fluctuations is the energy autocorrelation function 
\begin{equation}
C_{nm}(\varepsilon) = \int_{\Delta {\cal E}} {\rm d{\cal E}} (\sigma_{nm}({\cal E}+\varepsilon) - \bar{
\sigma}_{nm})( \sigma_{nm} ({\cal E})  - \bar{\sigma}_{nm}) \, .
\label{eq:-autocorr_inelastic}
\end{equation}
A least-square fit of the numerically obtained autocorrelation 
as depicted in Fig.~\ref{fig.Ericson} shows that it perfectly matches a Lorentzian 
\begin{equation}
\label{lorentz}
C_{nm}(\varepsilon) \propto {\Gamma^2\over \varepsilon ^2 + \Gamma^2} \, ,
\end{equation}
with mean resonance width $\Gamma=3.7\cdot10^{-3} J$ \footnote{We verified that a preparation in the regular regime
(i.e., $u\ll1$ or $u\gg1$), $C_{nm}(\varepsilon)$ shows significant deviations from a Lorentzian.}, which is one order of magnitude
larger than the mean level spacing $\Delta\approx 5 \cdot10^{-4} J$, directly extracted from our numerical data.
This is in perfect agreement with Ericson's scenario of overlapping resonances, and can be 
underpinned by a semiclassical picture \cite{BS88}:
\begin{figure}[t]
\centerline{\includegraphics[width=0.85\columnwidth,keepaspectratio]{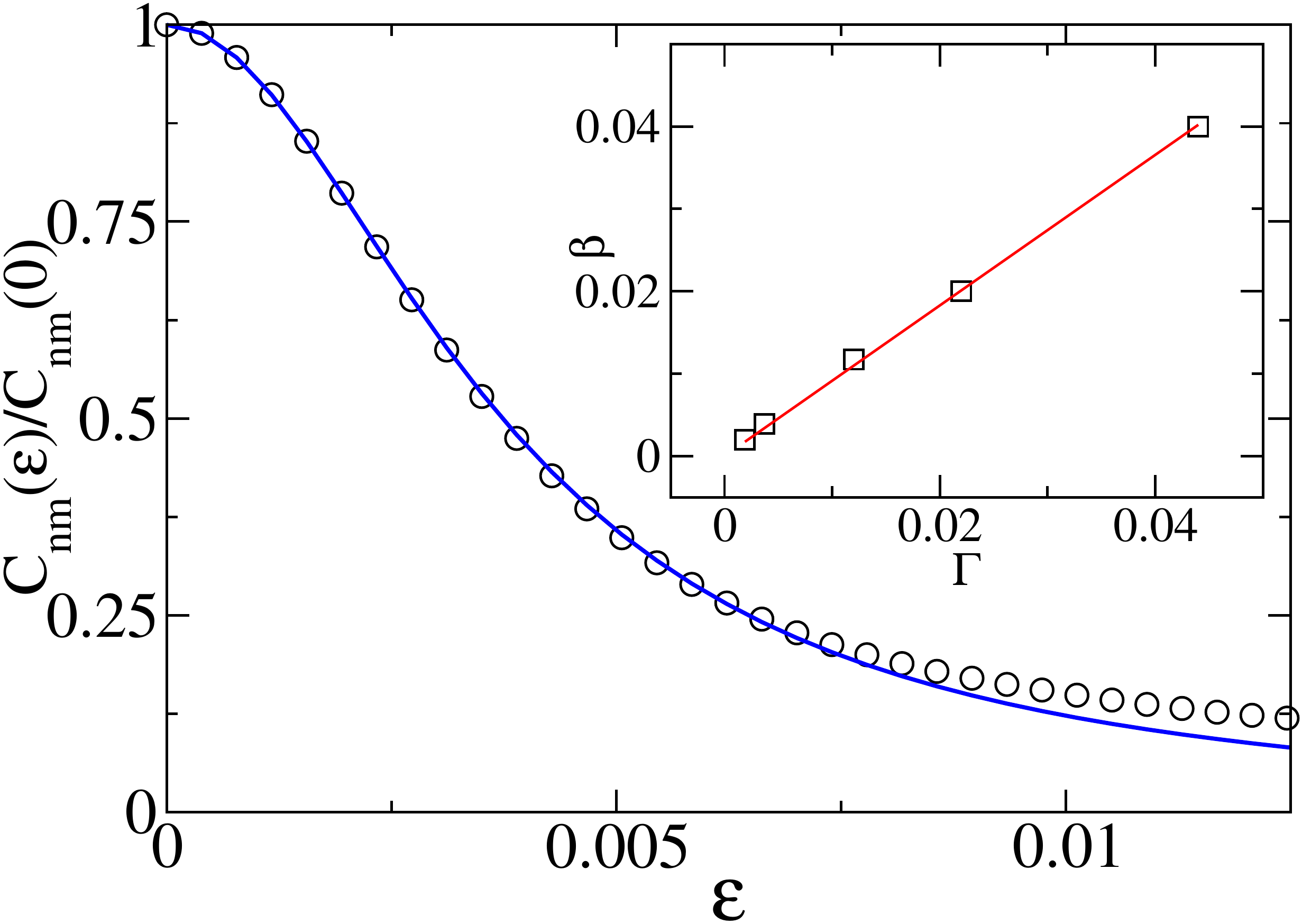}}
\caption{(color online). 
Autocorrelation function $C_{nm}(\epsilon)$ (black $\circ$) calculated from the inelastic scattering signal shown in 
Fig.~\ref{fig.rho_in}b).  The curve nicely matches a Lorentzian fit (blue).
{\it Inset:}
Semiclassical decay constant $\beta$ versus the mean resonance width $\Gamma$
(black $\Box$).
The data points correspond to different values of the coupling constant $\gamma$.
The semiclassical result is obtained after averaging over several initial conditions in the chaotic regime.
The predicted correspondence $(\beta=\Gamma)$ is confirmed by the fit $\beta= 0.92 \;  \Gamma$ (red).
} 
\label{fig.Ericson}
\end{figure}
The autocorrelation (\ref{lorentz}) can be interpreted as the squared  Fourier transform of the survival probability
$P(t)$ of the probe particle to stay a given time $t$ in the scattering region, i.e. on the TB site $j=0$,
hence with $P(t)=|c_0(t)|^2$. That latter quantity is evaluated by direct solution of the classical evolution equations
derived from (\ref{eq.GP}, \ref{eq.H_int}) (with initial conditions $P(0)=1$ and the GP system prepared at an energy
corresponding to $E_m$), and exhibits an exponential decay $P(t)\propto e^{-\beta t}$. $\beta$ thus determines the width 
of the (classical) autocorrelation function $C_{\rm clas} (\varepsilon)$, by virtue of 
\begin{equation}
\label{sp}
C_{\rm clas}(\varepsilon)= \left| \int dt \ P_{}(t) \exp(i\varepsilon t) \right|^2  \, ,
\end{equation}
which implies an average over all outgoing probe energies $\epsilon_n$. 
The inset of Fig.~\ref{fig.Ericson} demonstrates a perfect match of this semiclassically extracted
quantity $\beta$ with the width $\Gamma$ of the autocorrelation extracted from the quantum mechanical cross section,
and thus provides an independent, semiclassical proof of the Ericson scenario in the present many-particle problem.

\mysection{Conclusions}
In the light of recent BEC experiments, the proposed scattering setup represents an experimentally feasible and robust way
to non-destructively probe the condensate.
For example, a typical lattice depth of $V_0= 10\ E_{\rm R}$ recoil energies yields a tunneling strength $K=0.022$ $E_{\rm R}$ (cf. Eq.(20) of \cite{MO06}).
The exemplary case studied above ($u=5$, $N=38$) leads to a mean level spacing of $\Delta = 2.5\cdot10^{-3}$ $E_{\rm R}$.
This, in turn, corresponds to an energy resolution of $\Delta/E \approx 10^{-3}$ needed to resolve {\em single} scattering channels, i.e. the final state of the system.
As chaotic spectral statistics extend over a large parameter regime \cite{BK03, Venzl11} one could also choose larger lattices ($L>3$)
and/or fewer bosons to increase $\Delta$, and still obtain comparable results.
Let us also stress that, in the proposed measurement, it is not necessary to determine the boson number on the lattice:
As long as the detector can resolve the energy of the probe particle, the target properties can be inferred from the scattering signal,
without knowledge of the exact number of bosons in the system.

As expected, the observed chaotic scattering signals exhibit an inherent sensitivity on, e.g., the probe particle's energy.
To obtain a reliable characterization of the target, we employed a statistical analysis based mainly on the autocorrelation function $C(\varepsilon)$ and the probability distribution $P(\sigma_{ij})$ of the inelastic cross section.
On the other hand, the robustness of the obtained results against unavoidable fluctuations in the experimental control parameters is corroborated by averaging our results over $\Delta m=30$ channels (see, e.g., Fig.~\ref{fig.rho_in}).
This averaging, together with the statistical independence of the $S$-matrix elements, implies that our observations are robust against uncertainties, e.g., in the initially prepared state $|E_m\rangle$ of the condensate or in the particle number $N$, as well as in the resolution of the scattering channels.
For a typical laser wavelength of $1064~n{\rm m}$, $N=38$ Cs atoms, and $u=5$, this average corresponds to a residual
temperature of $\approx5~n{\rm K}$ what is readily achievable with state-of-the-art experiments that reach temperatures as low as $0.35~n{\rm K}$ \cite{Medl11}.

Hence, measurements of the partial inelastic cross section can identify an unambiguous case of Ericson fluctuations,
what yields robust information on the many-body spectrum. 
In contrast to compound nuclear reactions, here the latter is under perfect control, 
through the accurate experimental manipulation of the underlying Bose-Hubbard Hamiltonian, via the control parameter $u$.
This would allow to experimentally investigate the fate of the Ericson fluctuations at the transition from the chaotic to the regular regime.
By virtue of this control, our proposed setup could also add to the recent debate on complex many-body scattering \cite{Koeh10, Weid10,CAIZ11}.
Finally, the possibility to record single scattering channels could help the understanding of a related phenomenon:
Summing up an increasing number of contributing channels would then allow to study in a controlled way the crossover from the
Ericson regime to the multi-channel regime of universal conductance fluctuations.

\ack
We acknowledge financial support by DFG {\em Research Unit 760},
the US-Israel Binational Science Foundation (BSF), Jerusalem, Israel, a grant from
AFOSR No. FA 9550-10-1-0433, and a grant from the Deutsch-Israelische Projektkooperation (DIP).

\section*{References}

\bibliographystyle{iopart-num}
\bibliography{HHBCK_arxiv_12}

\end{document}